\documentclass{article}

\usepackage{comment}
\usepackage{fancyvrb}
\usepackage{code}
\usepackage{url}
\usepackage{graphics}
\usepackage{boxedminipage}

\begin{document}

\subsection*{Book review}
\emph{The Haskell Road to Logic, Maths and Programming},
Kees Doets and Jan van Eijck, Texts in Computing Series, 
King's College Publications, London, 2004. Price: \pounds 14.00 pounds, 
\$ 25.00, 444 pages, Paperback, ISBN: 0-9543006-9-6.

\bigskip
\noindent
Reviewer: Ralf L{\"a}mmel\\ Microsoft Corp., Redmond, U.S.A. 

\makeatactive

\section{Preamble}

In university curricula, the subjects of programming and maths and
logic tend to be separated. For instance, in the typical
computer-science curriculum, mathematics and logic are taught
according to the (dry) mathematical tradition, perhaps with modern
concessions to help the less abstract minds. One sort of concession is
to illustrate the implementability of selected mathematical concepts
in some programming language, be it C, Fortran, or others. Another
sort of concession is to hint at the usefulness of systems like Maple
or Mathematica~---~systems that are probably more appreciated by
mathematicians than applied, programming-minded computer
scientists. By contrast, programming courses are almost certainly
focused on a specific programming paradigm and on a specific
language. Logic and maths may show up in such courses as a sort of
``application domain'', while competing with many other domains.  For
instance, a typical course on the programming language Haskell is
likely to emphasize Haskell's strengths in the areas of abstract-data
type specification and combinator libraries for domain-specific
languages.

The authors of the book at hand have succeeded in amalgamating the
themes of maths, logic and Haskell programming in a mutually
beneficial manner. The book really stands out in so far that it
\emph{leverages} Haskell in diving deep into maths and logic (as deep as
covering power series or co-inductive proofs). The use of Haskell
makes the mathematical concepts quite digestible; Haskell lends itself
so well as a strongly typed and executable modeling language. The
authors' style is quite relaxing. The mathematical concepts are
explained in a way that carefully helps with an intuitive
understanding.

The ``Haskell road to Logic, Maths and Programming'' by Kees Doets and
Jan van Eijck does \emph{not} (mean to) comprise a detailed and
comprehensive introduction to functional programming or the
programming language Haskell. There exist such introductions to
Haskell; cf.\ the excellent textbooks by Thompson and
Hudak~\cite{Thompson96,Hudak00}, and these introductions should still
be useful for readers who want to deeply appreciate Haskell as a
general-purpose programming language and Haskell's strengths other
than maths and logic. The book at hand does not assume prior knowledge
of Haskell, though; all Haskell concepts are introduced incrementally.
The combination of texts on maths/logic with computer programming is
not unprecedented. For instance, Barwise and Etchemendy's ``Language,
Proof, and Logic''~\cite{BE99} accompanies a text on logic with
several computer programs. Even more to the point, Hall and
O'Donnell's ``Discrete mathematics using a computer''~\cite{HOD00}
covers maths and logic in similar ways than the book at hand, and it
also facilitates Haskell.


\section{Chapter by chapter}

\subsection*{Preface}

This is an outstanding book, but the preface could be improved.  In
particular, the preface seems to be somewhat ambiguous about the
purpose and the subject of the book. Citation: ``The purpose of the
book is to teach logic and mathematical reasoning in practice and to
connect logical reasoning with computer programming.'' The mentioning
of ``practice'' may count as confusing. Elsewhere in the preface:
``The subject of the book is the use of logic in practice, more in
particular the use of logic in reasoning about programming tasks.'' It
is not immediately obvious to the reader what sorts of ``programming
tasks'', what sorts of ``reasoning'' are meant here.  (For instance,
one may think that program validation and verification is included,
but this turns out not to be the case. Likewise, one may expect
reasoning about complexity.) It would help if the following
questions were answered more clearly: What are the
\emph{mathematical fields} covered by the book? What \emph{sorts
of reasoning} about either mathematical notions or programs or both
are carried out?

According to the preface, Haskell is a member of the LISP family with
the lambda calculus as foundation. This characterization does not hint
at two of Haskell's strengths: its advanced type system and its lazy
semantics. (LISP, in particular, does not share these assets.) Both
strengths are exploited in the book, and this is worth
emphasizing. The choice of Haskell is less arbitrary than it may seem
from the preface.

The outline of the chapters could be more helpful. Regarding chapters
1--7, there is basically one sentence. The explanations for the
remaining chapters may overwhelm newcomers with technical terms (cf.\
Chapter 9: ``closed forms for polynomial sequences''). It would be
wonderful to read a preface that provides an accessible overview of
the chapters, including a story how these subjects fit together,
making brief links to the three components of the book's title: logic,
maths, programming.

\subsection*{Chapter 1: Getting started}

The text starts with some basic explanations of literate programming,
lambda calculus and typed functional programming. Then, the text picks
up the problem of prime-number testing for a first illustration of the
``Haskell Road to Maths''. The text is pleasant to read, and it
introduces a good deal of mathematical conventions and Haskell idioms
including primitive types, products and lists. The text also states
clearly an important goal of the book early on: to communicate skills
for the formulation of properties about programs and their proofs. The
textbook style of the book also settles immediately: a very good mix
of definitions, explanations, examples and exercises.

Eventually, the chapter describes prime factorization and
prime-number generation. Surprisingly, the authoritative definition of
the former problem is given in an imperative language notation with
assignments! The Haskell implementation of the latter puts the
higher-order combinators @map@ and @filter@ to work. This is an
indication that the first chapter has come to real speed. Some bits of
the Haskell exposition in this chapter lack context in the sense that
idioms and examples do not contribute to the running theme of
algorithms around prime numbers.

The chapter pleasantly closes by emphasizing a major strength of
Haskell: this pure, declarative, functional language admits equational
reasoning, which is out of reach for imperative languages like Java
due to destructive assignments and other side effects in these
languages. Like most chapters in the book, Chapter 1 carries a short
tail ``Further Reading''.


\subsection*{Chapter 2: Talking about Mathematical Objects}

The chapter describes the basics of logic, in particular the usual
connectives (not, if, and, or, iff, for all, there exists) complete
with Haskell encodings of connectives, formulas, and validity checks
for formulas. The text is very pleasant to read; the alternation
between mathematical notation and Haskell encoding really makes
sense. The text carefully explains the many aspects of quantification:
free variables, open and closed formulas, domains for
quantification. The text makes a laudable effort to help the beginning
logician. For instance, the text identifies ``bad habits'' regarding
the ambiguous use of quantifiers in textual formulations of
mathematical statements.

The text makes clear that Platonism is adopted in this book: every
mathematical statement is either true or false. Other forms of logic,
such as intuitionistic logic or fuzzy logic are not covered.

\subsubsection*{A side note on Haskell's type classes}

Throughout the book, there are a few occasions where Haskell's type
classes could show off nicely. For instance, when checks for validity
or logical equivalence are provided (cf.\ Chapter 2), they are coded
separately for \emph{several} arities, as indicated by the following
examples:

\begin{verbatim}
 valid1 :: (Bool -> Bool) -> Bool
 valid1 bf =  (bf True)
           && (bf False)
\end{verbatim}

\newpage

\begin{verbatim}
 valid2 :: (Bool -> Bool -> Bool)  -> Bool
 valid2 bf =  (bf True  True)
           && (bf True  False) 
           && (bf False True)
           && (bf False False)

 valid3 :: (Bool -> Bool -> Bool -> Bool) -> Bool
 -- ... and so on ...
\end{verbatim}

A single, generic @valid@ function would suffice for all arities, as
defined below. The same holds for logical equivalence. (It is left as
an exercise for the reader to eliminate the minor violation of strict
Haskell 98 rules.)

\begin{verbatim}
 class Valid f
  where
   valid :: f -> Bool

 instance Valid Bool
  where
   valid = id

 instance Valid f => Valid (Bool -> f)
  where
   valid f =  valid (f True)
           && valid (f False)
\end{verbatim}

It is understandable that the text does not want to overwhelm the
reader in the early chapters. However, the power of Haskell's type
classes could have been discussed in a later chapter. Fortunately, 
the reader will still hear about this important part of Haskell
because a few chapters take advantage of some type classes from 
the Haskell library.


\subsection*{Chapter 3: The Use of Logic: Proof}

The chapter communicates style recommendations and recipes for simple
proofs. This is done without much reference to Haskell or Haskell
examples. The idea seems to be that the chapter provides the
foundation for carrying out mathematical proofs and proofs about
program properties in all subsequent chapters.

A typical style recommendation is ``Make sure the reader knows exactly
what you are up to.'' The various recipes basically cover all logical
connectives~---~their introduction and elimination. And then again,
even the use of the recipes is complemented by further
recommendations such as not to consider \emph{proof by
contradiction} all too easily.

The authors deserve extra credit for their clear and detailed
enumeration and rationalization of the proof recipes, complete with
careful coverage of style issues. This textbook may really enable
readers to acquire proving skills. Some of the examples may be
potentially too demanding for the beginning mathematician. For
instance, the first example for $\exists$-elimination refers to
transcendent reals and their definition as not being roots of a
polynomial with integer coefficients. (Polynomials and number theory
are only covered much later in the book.)

The text readily admits that programming languages like Haskell are of
little help with proving general properties about programs. One may
use the computer though to refute single cases.  Perhaps, this is too
much to ask for, but two topics could have been touched upon in the
context of this chapter:
\begin{itemize}
\item The Curry-Howard isomorphism, i.e., the correspondence between
types and logical formulas, as much as expressions and logical proofs;
see~\cite{Thompson91} for a functional-programming-minded discussion
of this correspondence.
\item The use of theorem provers to check or even to find proofs. 
\end{itemize}


\subsection*{Chapter 4: Sets, Types and Lists}

The chapter, according to the authors, presents ample opportunity to
exercise (perhaps better: acquire) the reader's skills in writing
implementations for set operations and for proving things about sets.
This is perhaps a bit misleading. The text does certainly not discuss
different (efficient) implementations of sets or other data types. In
reality, the chapter focuses on the axiomatization of sets as well as
pairs, products and lists. The text also explains Russell's paradox
and the provisions by the Haskell type system that help avoiding such
problems in actual Haskell programs. (The text also touches upon the
relation between the Halting problem and Russell's paradox.) 
Eventually, several list- and set-processing primitives are
encoded. These implementations facilitate the type classes @Eq@
(``equality'') and @Ord@ (``ordering'') from the Haskell
prelude. Afterwards, the chapter spends some time on Haskell's special
idioms for list comprehensions, which are clearly related to the
mathematical notion of set comprehensions. List comprehensions are
illustrated with a detailed example in the realm of relational
database queries.

Regarding ``Further Reading'', some more references spring to mind.
The reader could appreciate pointers to existing
functional-programming-minded or Haskell-specific work on the
efficient implementation of data types~\cite{Okasaki96,Edison}. The
chapter's major example for list comprehensions, a database query
scenario, is weakly typed: the type @String@ is used as the universal
data type for all cells in the database instance and all column names.
Haskell can do much better using more sophisticated type-system
features. Admittedly, substantially more strongly typed approaches
require non-Haskell~98 features~\cite{LM99,HList}. It is an asset of
the book at hand that it stays with the Haskell~98 standard.


\subsection*{Chapter 5: Relations}

The chapter describes the theory of relations (``sets of ordered
pairs''); it defines the common properties of relations and other
related terms (reflexivity, transitivity, equivalence relations,
quotients, characteristic functions, etc.). The properties are also
implemented as checks that can be effectively computed for relations
that are given as finite sets of pairs. The chapter focuses on
elementary material about relations. N-ary relations and the
corresponding relational algebra are not covered. The chapter offers a
huge amount of simple but insightful exercises with which the reader
can develop skills in mathematical proofs. These exercises are
particularly manageable since they assume mostly basic properties of
sets and the chapter's simple definitions for relations.


\subsection*{Chapter 6: Functions}

The chapter is about functions as \emph{mathematical objects}, not to
be confused with Haskell functions. There is clearly a plethora of
important mathematical notions around functions: function definition
by case discrimination, the relational view on functions, surjections,
injections, bijections, inverse, partiality, congruences, and so on.

As in previous chapters, the text captures properties of mathematical
objects as Haskell expressions that perform ``checks''.  That is, a
certain property of a function, such as injectivity, is checked under
the assumption that the domain of the function is given as a (finite)
list:

\begin{verbatim}
 injective :: Eq b => (a -> b) -> [a] ->  Bool
 injective f [] = True
 injective f (x:xs) =
     notElem (f x) (image f xs) && injective f xs
\end{verbatim}

One could take this idea a tiny step further, by writing down
QuickCheck~\cite{CH00} properties in Haskell. The QuickCheck tool
would automatically check all these properties. For instance, rather
than stating the mere expectation that @fromEnum@ should be the
left-inverse of @toEnum@ (cf.\ page 222), one could instead formulate
QuickCheck properties for different instances of @Enum@ (i.e., for
different enumeration types).  Here is a sample property for the
@Char@ type:

\begin{verbatim}
 prop_EnumChar = forAll charInts $ \x -> 
                   fromEnum ((toEnum x) :: Char) == x
 charInts :: Gen Int
 -- restrict Int generator to valid character codes
\end{verbatim}


\subsection*{Chapter 7: Induction and Recursion}

The chapter starts with the basic proof method for mathematical
induction. The method is explained in all detail and with great
care. In particular, the text clarifies that induction proofs are
particularly straightforward whenever the function ``under study'' is
defined by (primitive) recursion on the natural numbers. Induction and
recursion is then also demonstrated for trees and lists. The text
clearly hints at the full generality of the concepts: recursion
schemes can be advised for a large class of data types, and
corresponding induction proof schemes are thereby enabled.

The ``Further Reading'' section would be more complete with some
references to seminal work on recursion schemes,
e.g.,~\cite{MFP91}. There is also a wonderful tutorial by Augusteijn,
which applies morphisms (such as primitive recursion) to the rigorous
design of sorting algorithms and their encoding in
Haskell~\cite{Augusteijn98}.


\subsection*{Chapter 8: Working with Numbers}

The chapter presents the different number systems that exist: natural
numbers, integers, rational numbers, irrational numbers, and complex
numbers. By developing these number systems for Haskell, the text also
provides deeper insight into Haskell's built-in and library support
for numbers. 

As usual, natural numbers are shown to suffice as basis for
integers and rational numbers. In fact, different representations for
these two derived number systems are discussed and implemented in
Haskell. Irrational numbers are defined informally (as infinite,
non-periodic decimal expansions). Implementation-wise, the Haskell
primitive type for reals is used. Key notions for reasoning about
reals are defined: continuous functions, limits, Cauchy sequences.

Throughout the chapter, it is shown how the different number systems
participate in standard type classes for equality, ordering and
arithmetic operations. Standard algebraic properties of the operations
on the numbers (associativity etc.) are routinely proved. For several
of the number representations, geometric interpretations are
visualized. These illustrations should be of great help for the
reader.


\subsection*{Chapter 9: Polynomials}

The chapter discusses polynomials with the initial goal to automate
the search for polynomials as closed forms for sequences of
integers. To this end, Babbage's classic difference method is
explained and implemented in Haskell. (Recall that the method works as
follows: suppose we are given a ``long enough'' sequence of integers
$a_0$, \ldots, $a_n$, for which we presume that it may be computed
from a polynomial $f(x)$ such that $a_0 = f(0)$, \ldots, $a_n = f(n)$,
then we may attempt to find the closed form $f(x)$ in two steps. (i)
Determine the degree of $f(x)$ by difference analysis on the given
sequence of integers. (ii) Perform Gaussian elimination to determine
the coefficients of $f(x)$.)

The authors succeed in capturing the beauty of this method, making
clear that no magic is involved. Again, the combination of
mathematical notation and Haskell encoding helps enormously to get all
the fine details across. This chapter is a master piece: a concise but
still comprehensive and insightful explanation of the method. All of
Babbage, complete with the Haskell encoding on just 13 pages!

The chapter also discusses the link between polynomials and
combinatorics. In this context, Newton's binomial theorem is derived
and put to work. These parts of the chapter are slightly less clear in
stating \emph{up-front} what they aim to achieve.

We recall the simple version of Newton's binomial theorem:

\[(z+1)^n = \sum_{k=0}^{n}(\begin{array}{l}n\\k\end{array}) z^k\]

The left-hand side is not in the form of a polynomial (it uses a
binomial~---~a sum of two terms), whereas the right-hand side is in
the form of a polynomial. Hence, one way to read Newton's theorem is
that it enables binomial expansions such that pure polynomials are
derived.

It is straightforward to prove Newton's theorem (by induction). The
text however conveys the beauty of Newton's findings by deriving the
theorem through operations on polynomials: substitution and
calculus. Again, the text communicates deep insight combined with
a Haskell model.

The final part of the chapter discusses the representation of
polynomials as lists of their coefficients. Polynomial arithmetics and
calculus is then derived on top of this representation. The main goal
is to reason about combinatorial problems.  In particular, polynomials
(as lists of coefficients) are derived as solutions to combinatorial
problems. For instance, the polynomial expansion of $(z+1)^{10}$
solves the problem of picking $k$ elements from a set of 10.


\subsection*{Chapter 10: Corecursion}

The chapter discusses infinite data structures (mostly streams).
Haskell laziness's makes this language clearly suited to study such
data structures even programmatically. One important goal of the
chapter is to describe proof methods for corecursive functionality.
The text also makes excursions to the following two related subjects.
Non-deterministic processes are modeled as functions from random
integer streams (for coding ``decisions'') to streams (for coding the
actions performed). The earlier discussion of the link between
polynomials and combinatorics is generalized to power series (so to
say polynomials with infinite series of coefficients). Power series
naturally show up once polynomials are divided; the results of
division may have an infinite degree.

The authors make the point (in the introduction) that they deliver the
first, general textbook treatment of recursion. This already indicates
the challenge of this section. Some readers may expect from the
chapter to become well-versed in coinduction proofs. Unfortunately,
the text is scarce in this respect. There are essentially just 5 pages
for the recipe of coinduction proofs. There is essentially just one
example of a coinduction proof for streams, and one may argue whether
or not the proof is detailed enough. The ``Further reading'' section
should perhaps refer to more scenarios for coinductive reasoning, and
more sample proofs.  In particular, Barwise and Moss' book on
``Vicious circles'' ~\cite{BM96} springs to mind.


\subsection*{Chapter 11: Finite and Infinite Sets}

The chapter can be seen as a more advanced continuation of Chapter 4:
Sets, Types and Lists. The text begins with a deeper analysis of
mathematical induction. Many additional insights are provided, when
compared to the intuitive use of induction proofs in earlier
chapters. For instance, the notion of strong induction is presented,
which does not require a base step and uses a stronger induction
hypothesis, thereby enabling proofs of more involved problems. Another
subject of the chapter is equipollence, preparing ultimately for the
discussion of cardinalities of sets including infinite cardinalities
such as $\aleph_0$ (i.e., the cardinality of the natural numbers) and
$2^{\aleph^0}$ (i.e., the cardinality of the powerset of the natural
numbers). A few ideas are nicely illustrated through Haskell code,
e.g., the diagonalization for the rational numbers, which shows that
this set of numbers is countably infinite.


\section{Non-Haskell roads to logic and maths}

When reading this book, it is a useful exercise to question every now
and then whether or not the Haskell road is a convenient one, what the
issues would be when mainstream programming languages were used at
times.

It seems to be undisputed that Haskell allows for a more or less
immediate transcription of mathematical notation to program code. This
is made possible by Haskell's declarativeness, purity and
conciseness. Haskell's laziness further helps with encoding
mathematical problems that involve infinite data structures. Haskell's
higher-orderness is essential to provide mathematical concepts as
library functionality. (For instance, think of the higher-order
function that turns a sequence of a polynomial's coefficients in the
actual function for the polynomial.) Finally, Haskell's type system
serves an important documentary purpose; machine-checked type
information and executability are in fact the two major ways in which
Haskell notation \emph{complements} mathematical notation.

Without any claim of completeness, we would like to illustrate some of
the issues that arise when encoding maths in mainstream languages such
as C\# or Java. For the code examples that follow, we will use C\#
(2.0), but we could have used Java (1.5) just as well, without
changing the tenor of the observations.

Consider the following Haskell function, @difs@, for computing a
difference sequence. This operation is at the heart of Babbage's
difference engine. (Recall: one uses the function @difs@ to determine
the degree of a potential polynomial $f(x)$ that should be derived as
the closed form from a ``long enough'' sequence of integers $a_0$,
\ldots, $a_n$, for which we assume that it corresponds to $f(0)$,
\ldots, $f(n)$. The degree of $f(x)$ is the number of iterations of
applying @difs@ until the resulting sequence is a constant sequence,
if this happens at all.)

\bigskip

\noindent
\begin{boxedminipage}{\hsize}
\begin{verbatim}
 difs :: [Integer] -> [Integer]
 difs []  = []
 difs [_] = []
 difs (m:n:ks) = n-m : difs (n:ks)
\end{verbatim}
\end{boxedminipage}

\bigskip

\noindent
The following C\# encoding may serve as an efficient transcription.

\bigskip

\noindent
\begin{boxedminipage}{\hsize}\small
\begin{verbatim}
 static Decimal[] difs(Decimal[] given)
 {
    Decimal[] result = new Decimal[
      given.Length == 0 ? 0 : given.Length-1];
    for (int i=1; i<given.Length; i++)
      result[i-1] = given[i]-given[i-1];
    return result;
 }
\end{verbatim}
\end{boxedminipage}

\bigskip

The C\# function operates on arrays of values of type @Decimal@ (which
is C\#'s counterpart to Haskell's arbitrary precision integers). The
array for the result is explicitly allocated, leveraging the
observation that the @difs@ function returns a sequence that is one
element shorter than the (non-empty) input sequence. All sense of
equational reasoning is gone. The allocation code requires an extra
proof in order to establish its correctness. There is actually no
straightforward way to retain the recursive formulation of the
problem. (For the record, with C++ pointer arithmetics, we were able
to process the input array recursively. However, the recursive
synthesis of the result would not work so easily.)

Hence, the mainstream road to maths is bumpy because of the encoding
efforts; it will be rather painful once we wanted to reason about
these programs. We attempt some variations in the sequel. The
following C\# variation uses the generic collection type @List@ in
place of arrays:

\bigskip

\noindent
\begin{boxedminipage}{\hsize}\small
\begin{verbatim}
 static List<Decimal> difs(List<Decimal> given)
 {
    List<Decimal> result = new List<Decimal>();
    if (given.Count > 1)
    {
       result.Add(given[1]-given[0]);
       result.AddRange(difs(given.GetRange(1,given.Count-1)));
    }
    return result;
 }
\end{verbatim}
\end{boxedminipage}

\bigskip

The earlier array location reduces to a simple object construction for
the initially empty list. (Hence, we eliminated the need for the extra
insight regarding the length of the result.) The @List@ type is rich
enough to re-establish the recursive coding style. The exercised style
is rather inefficient. For instance, the @GetRange@ method involves
cloning of the ``tail''. We also note that we still fail to provide
any sort of equational style; so equational reasoning remains out of
reach. Furthermore, we fail to obtain the conciseness of pattern
matching.

Yet another deficiency is worth mentioning: The previous C\# encodings
are all eager. This is perhaps acceptable for the use of @difs@ in the
context of Babbage's method, but eventually all sorts of similar list
processing functionality needs to be lazy. (Think of moving from
polynomials to power series.) Here is a more clever C\# solution,
which engages in less cloning, and which is lazy at the same time:

\bigskip

\noindent
\begin{boxedminipage}{\hsize}\small
\begin{verbatim}
 static IEnumerable<Decimal> difs(IEnumerable<Decimal> given)
 {
    Decimal? i = null;
    foreach (Decimal j in given)
    {
       if (i.HasValue) yield return j-i.Value;
       i = j;
    }
 }
\end{verbatim}
\end{boxedminipage}

\bigskip

That is, we use the @IEnumerable@ interface and the @yield@ @return@
idiom to compute the resulting list lazily. (We also use C\#'s
nullable types; cf.\ ``?''.) It is not an accident that this solution
uses an imperative loop again. That is, C\#'s streams (or lazy lists)
are idiomatically restricted. Re-establishing recursion is a
non-trivial undertaking. Here is a (prohibitively inefficient) attempt:

\bigskip

\noindent
\begin{boxedminipage}{\hsize}\small
\begin{verbatim}
 static IEnumerable<Decimal> difs(IEnumerable<Decimal> given)
 {
    if (empty(given)) yield break;
    if (empty(tail(given))) yield break;
    Decimal m = head(given);
    Decimal n = head(tail(given));
    yield return n-m;
    foreach (Decimal i in difs(tail(given)))
      yield return i;
 }
\end{verbatim}
\end{boxedminipage}

\bigskip

One major problem is that the streams returned from the recursive
calls needs to be pulled into the local result stream explicitly
through the (expensive) @foreach@ loop. Also, the code is baroque
anyhow, even though we anticipated the possibility of helpers to take
apart lists; cf.\ @empty@, @head@, @tail@ (definitions omitted).

\paragraph*{Bottom line}

This detour has illustrated that relatively simple pieces of
mathematical definitions can be difficult to encode in mainstream
programming languages. So anyone in search of a road to logic and
maths, is well advised to favor the Haskell road, not to get lost in
the programming bits of such an endeavor. It is a unique capability of
Haskell that mathematical notation (in the form of equations, sums,
products, compound expressions, etc.) is type-able and executable in
this programming language more or less as is. The book under review
leverages this capability very appropriately.


\newpage

\section{Concluding remarks}

Doets and van Eijck's ``The Haskell Road to Logic, Maths and
Programming'' is an astonishingly extensive and accessible textbook on
logic, maths, and Haskell. The book adopts a systematic but relaxed
mathematical style (definition, example, exercise, ...); the text is
very pleasant to read due to a small amount of anecdotal information,
and due to the fact that definitions are fluently integrated in the
running text. The overall selection of mathematical subjects is
non-surprising (which is good), except that some advanced topics are
included such as corecursion and Cantor's infinities. One would
perhaps expect to see a more substantial treatment of calculus and
some coverage of statistics.

Regarding the targeted audience, the text seems to be valuable and
advisable for undergraduates with mathematics and logic as minor
subjects. However, even mathematically ambitious readers should find
this book interesting, be it then more for the Haskell part of it. The
book was indeed developed through undergraduate courses at the
University of Amsterdam.

The Haskell samples in the book appear to be highly idiomatic, and
readers can certainly expect to learn Haskell programming, so to say
as a side effect. Not surprisingly, the book does not cover all of a
typical functional programming text. For instance, type classes are
used only in a minimalistic way. Also, monads do not show up anywhere
in the book. While this is normally ``unimaginable'' for a Haskell
introduction, it makes total sense for this specific book.  The
book's write-up and program development has been based on literate
programming. (However, the sources that are obtainable from the
authors' website do not currently comprise the text of the chapters;
presumably for copyright reasons.)

Regarding the availability of the book: the ``Texts in Computing
Series'' of \emph{King's College Publications} are published on a
"Print on Demand" basis. This has as the benefit that the price is
much lower than with regular publishers~---~at the time of writing:
25\$ vs. 14\pounds. Some may argue that it also has the minor drawback
that the book is not for sale in local bookstore, but has to be
ordered through Amazon instead.

I take the liberty to close this review with an admittedly personal
comment. I would have wanted this very book during my computer science
studies. The Haskell way of modeling mathematical concepts in a
declarative, executable and strongly typed manner is very supportive
in learning. The small but telling distance between mathematical
notation and Haskell notation makes it also more rewarding to actually
engage in reasoning about mathematical problems and programs. I still
remember wasting time on engaging in Pascal encodings for some of my
maths homework. Should I end up giving lectures on maths and logic, I
will love to use this book in such a course.


\newpage

\bibliography{paper}
\bibliographystyle{abbrv}

\end{document}